\documentclass[12pt]{article}

\usepackage{amsmath}
\usepackage{amssymb}
\usepackage{cite}

\def\nn{\nonumber}

\numberwithin{equation}{section}

\title{\bf \Large Lagrangian formulation for perfect fluid equations\\
with the $\ell$-conformal Galilei symmetry }

\author{Timofei  Snegirev${}^{a}$\thanks{timofei.v.snegirev@tusur.ru}
\\[0.5cm]
\it{\small ${}^a$Laboratory of Applied Mathematics and Theoretical Physics,}\\
\it{\small Tomsk State University of Control Systems and Radioelectronics,}\\
\it{\small Lenin ave. 40, 634050 Tomsk, Russia}}

\date{}

\begin{document}

\maketitle

\begin{abstract}
Lagrangian formulation for perfect fluid equations which hold
invariant under the $\ell$-conformal Galilei group
with half-integer $\ell$ is proposed. It is based on a
Clebsch-type parametrization and reproduces Lagrangian description
of the Euler fluid equations for $\ell=\frac12$. The transition from the
Lagrangian formulation to the Hamiltonian one is analyzed in
detail.

\end{abstract}

\thispagestyle{empty}
\newpage
\setcounter{page}{1}

\section{Introduction}\label{Sec1}

Fluid mechanics with conformal symmetries currently attracts
considerable attention in connection with the
AdS/CFT-correspondence \cite{Mal97} and the flud/gravity duality
\cite{Ran09}. In particular, the latter can be understood as a
hydrodynamic limit of the former in which the formalism of fluid mechanics
is applied with the aim to an effective description of a strongly coupled
quantum field theory. At the same time successful efforts to extend
holography to strongly coupled condensed matter systems
\cite{Son08,BM08,NS2010} stimulate investigations of fluid dynamics
with non-relativistic conformal symmetries.

In contrast to the unique relativistic conformal algebra, there are
several options available in the non-relativistic case. A well known
example is the Schr\"odinger algebra \cite{Jack72,Nied72,Hag72},
which has been found to be relevant for a wide range of physical
applications (see the review \cite{DHHR24} and references therein).
The Schrodinger group was originally discovered as the maximal
kinematical invariance group of the Schrodinger equation for a free
massive particle~\cite{Nied72}\footnote{In fact, similar
non-relativistic conformal structure has been known since 19th
century due to the work on classical mechanics\cite{Jac1884} and the
heat equation \cite{Lie1881}.}. In addition to the Galilei
transformations it contains dilatation and special conformal
transformation. Surprisingly enough, the non-relativistic
contraction of the relativistic conformal algebra \cite{LSZ05} does
not result in the Schrodinger algebra. The latter fact stimulates
interest in the study of other finite-dimensional conformal
extensions of the Galilei algebra which are combined into a family
known in the literature as the $\ell$-conformal Galilei algebra
\cite{Henkel,NOR97}.

The algebra is characterized by an arbitrary integer or half-integer
parameter $\ell$ and, in addition to temporal translation,
dilatation and special conformal transformation, it involves a set
of vector generators $C_i^{(k)}$, where $i=1, \dots, d$ is a spatial
index and $k=0,...,2\ell$. $C_i^{(0)}$ and $C_i^{(1)}$ link to
spatial translations and Galilei boosts while higher values of $k$
correspond to the so called constant accelerations. The case
$\ell=\frac12$ reproduces the Schrodinger algebra, while $\ell=1$ is
recovered in the non-relativistic limit of the relativistic
conformal algebra. The latter is usually referred to as the
conformal Galilei algebra \cite{LSZ05}.

As far as dynamical realizations of the $\ell$-conformal Galilei
group are concerned (see e.g.
\cite{LSZ1,DH,FIL,DH1,GK,GM3,AGKM,AGGM,CG} and references therein),
the physical meaning of the parameter $\ell$ may vary. In condensed
matter physics, the reciprocal $z=1/\ell$ is known as a critical
dynamical exponent, which links to the fact that under dilatation
temporal and spatial coordinates scale differently, $t'=\lambda t$,
$x'_i=\lambda^\ell x_i$, i.e. $\ell$ characterizes the degree of
scaling anisotropy. In mechanics, field theory, and fluid mechanics,
$\ell$ determines the order of differential equations of motion.
Because a number of functionally independent integrals of motion
needed to integrate a differential equation correlates with its
order, in order to accommodate symmetries generated by the tower of
vector generators $C_i^{(k)}$, one has to make recourse to higher
derivative systems. In particular, in mechanics and field theory the
order of differential equations of motion is $2\ell+1$, while in
fluid mechanics it is $2\ell$. To give a notable example, the
celebrated Pais-Uhlenbeck oscillator \cite{PU} enjoys the
$\ell$-conformal Galilei symmetry for a special choice of its
frequencies \cite{AGGM}.

Previous studies of non-relativistic conformal symmetries in the
context of fluid mechanics revealed interesting results. A perfect
fluid described by the Euler equations has the Schr\"odinger
symmetry ($\ell=\frac12$) provided a specific equation of state is
chosen \cite{HH1,RS00} which links pressure to density. For a
viscous fluid the Schrodinger symmetries are partially broken
leaving one with the dilatation and Galilei symmetries \cite{HZ}.
Attempts to discover the conformal Galilei symmetry ($\ell=1$) for
systems which derive from relativistic conformally invariant
hydrodynamic equations did not lead to success
\cite{FOz08,FOz09,HZ}. In the non-relativistic limit such systems
proved to be of limited physical interest. Reasonable hydrodynamic
equations can be obtained as a result of a more subtle
non-relativistic contraction, but they do not enjoy conformal
Galilei  symmetries. It was recently shown that one can construct
generalized perfect fluid equations which accommodate the
$\ell$-conformal Galilei symmetries for an arbitrary $\ell$
\cite{Gal22a,Gal22b}. In particular, these equations contain the
generalized Euler equation with higher material derivatives, which
reduces to the perfect fluid equations for $\ell=\frac12$.

Given a set of equations of motion, it is always desirable to have a
Lagrangian formulation. There are two approaches to describe the
non-relativistic perfect fluid
equations\cite{Ser59,Eck60,AK88,JNPP04,BJLNP02} (for modern
developments see also \cite{DHNS11,NRR11,CNT13,Nai20,Tor23}). The
first approach (the Lagrange picture) deals with coordinates of a
fluid particle parametrized by a set of continuum labels (see the
classic monograph \cite{AK88}). In this case, the Lagrangian
function has the same form as in Newtonian mechanics and the action
is invariant under volume-preserving diffeomorphism. The second
approach (the Euler picture) is akin to classical field theory and
deals directly with physical variables which are interpreted as the
fluid density and the velocity vector field. Here the fluid
equations can be naturally put into the Hamiltonian form
\cite{MG80}. The peculiarity of the latter formulation is that the
Poisson brackets among the physical fields are non-canonical. In
order to identify canonical variables and go over to a Lagrangian
description, the Clebsch parametrization \cite{Cleb1859} of the
velocity vector field is usually used. The latter description as
well as a relation between the two approaches to fluid mechanics are
reviewed in detail in \cite{JNPP04}.

It was recently shown \cite{Sne23a} that
the approach involving non-canonical Poisson brackets \cite{MG80}
can be adapted to construct a Hamiltonian formulation of the
generalized perfect fluid equations with the $\ell$-conformal
Galilei symmetries for an arbitrary half-integer $\ell$. Aiming at a
Lagrangian formulation, it is important to understand whether
canonical variables exist in which the perfect fluid equations with
the $\ell$-conformal Galilei symmetries arise from the variational
principle. The goal of this work is to elaborate on this issue.

The paper is organized as follows. In the next section, we briefly
review the generalized perfect fluid equations invariant under the
$\ell$-conformal Galilei group \cite{Gal22a} and their Hamiltonian
formulation developed in \cite{Sne23a}. In Sect. \ref{Sec3}, we
construct the Lagrangian formulation based on a Clebsch-type
parametrization in which the generalized perfect fluid equations
arise from the variational principle. In Sect. \ref{Sec4}, the Dirac
method is used to analyze constraints which arise after transition
to the Hamiltonian formalism. A relation to the Hamiltonian
description in terms of non-canonical Poisson brackets \cite{Sne23a}
is studied in detail. In the concluding Sect. \ref{Sec5}, we
summaries our results discuss possible further developments.

\section{{Equations of motion  and Hamiltonian formulation}}\label{Sec2}

Let us take briefly remind the
structure of the $\ell$-conformal Galilei  algebra \cite{NOR97}. Its generators
include temporal translation $H$,
dilatation $D$, special conformal transformation $K$ and a set of
vector generators $C_i^{(k)}$, $k=0,...,2\ell$. The latter
correspond to spatial translations ($k=0$), Galilei boosts ($k=1$) and
the so-called constant accelerations ($k>1$). The structure
relations of the algebra read\footnote{The algebra also includes spatial rotation which in
what follows will be disregarded.}
\begin{align}\label{lCG}
& {[H,D]}=H, && {[H,C^{(k)}_i]}=kC^{(k-1)}_i,\nn
\\
& {[H,K]}=2D, && {[D,C^{(k)}_i]}=(k-\ell)C^{(k)}_i,\nn
\\
& {[D,K]}=K, && {[K,C^{(k)}_i]}={(k-2\ell)}C^{(k+1)}_i.
\end{align}
They can be realized in a non-relativistic space-time parameterized
by $(t,x_i)$, $i=1,...,d$, by the following way\footnote{Throughout
the text we use the notations: $\partial_0=\frac{\partial}{\partial
t}$, $\partial_i=\frac{\partial}{\partial x_i}$. Summation over
repeated indices is understood.} \cite{NOR97}
\begin{eqnarray*}
H={\partial}_0,\quad D=t{\partial}_0+\ell x_i{\partial}_i,\quad
K=t^2{\partial}_0+2\ell tx_i{\partial}_i,\quad
C^{(k)}_i=t^k{\partial}_i.
\end{eqnarray*}

Generalized perfect fluid equations invariant under the
$\ell$-conformal Galilei group were formulated in a recent paper
\cite{Gal22a}
\begin{eqnarray}\label{PFEqdl}
{\partial_0\rho}+ {\partial_i (\rho\upsilon_i)}=0,\quad {\cal
D}^{2\ell}\upsilon_i=-\frac{1}{\rho}{\partial_i p},\quad
p=\nu\rho^{1+\frac{1}{\ell d}},
\end{eqnarray}
where $\rho(t,x)$, $\upsilon_i(t,x)$, $p(t,x)$ are the
density, the velocity vector field, and the pressure, respectively, and ${\cal
D}=\partial_0+\upsilon_i\partial_i$ is the material derivative. The
first equation is the continuity equation, while the second and third
equations describe the generalized Euler equation with higher
derivatives and the equation of state which links the pressure to the
density, $\nu$ being a constant. For $\ell=\frac12$, the
equations (\ref{PFEqdl}) reproduce the perfect fluid equations
invariant under the action of the Schrodinger group \cite{RS00}. In what follows, we
will refer to the model (\ref{PFEqdl}) as the
$\ell$-conformal perfect fluid.

For half-integer values $\ell=n+\frac12,\; n=0,1,...$, the equations
(\ref{PFEqdl}) admit a Hamiltonian formulation \cite{Sne23a}. In order to construct it,
auxiliary fields $\upsilon^{(k)}_i$,
$k=0,1,...,2n$ are introduced with $\upsilon^{(0)}_i=\upsilon_i$ and the second
equation in (\ref{PFEqdl}) is rewritten as the equivalent first order
system
\begin{eqnarray}\label{HamFlEql}
{\cal D}\upsilon^{(k)}_i=\upsilon^{(k+1)}_i ,\quad {\cal
D}\upsilon^{(2n)}_i=-\frac{1}{\rho}{\partial_i p}.
\end{eqnarray}
Then one can verify that the following Hamiltonian
\begin{eqnarray}\label{Hamdl}
H&=&\int dx
\left(\frac12\rho\sum_{k=0}^{2n}(-1)^k\upsilon^{(k)}_i\upsilon^{(2n-k)}_i+V\right),\quad
V=\ell d p,
\end{eqnarray}
puts the original equations of motion into the Hamiltonian form
$\partial_0\rho=\{\rho,H\}$,
$\partial_0\upsilon_i^{(k)}=\{\upsilon_i^{(k)},H\}$ provided the non-canonical Poisson brackets
\begin{eqnarray}\label{PBPFl}
\{\rho(x),\upsilon^{(k)}_i(y)\}&=&-\delta_{(k)(2n)}{\partial_i}\delta(x-y),\nn
\\
\{\upsilon^{(k)}_i(x),\upsilon^{(m)}_j(y)\}&=&\frac{1}{\rho}\left(\delta_{(k)(2n)}{\partial_i
\upsilon^{(m)}_j}- \delta_{(m)(2n)}{\partial_j
\upsilon^{(k)}_i}\right.\\
&&\left.\phantom{\upsilon^{(m)}_j}+(-1)^{k+1}\delta_{(k+m)(2n-1)}\delta_{ij}\right)\delta(x-y)\nn
\end{eqnarray}
are introduced.
Here $\delta_{(k)(m)}$ is the Kronecker symbol.

Within the Hamiltonian formalism the $\ell$-conformal
Galilei algebra is realized as follows. The Hamiltonian $H$
(the conserved energy) (\ref{Hamdl}) links to temporal translation
while conserved charges associated with the dilatation, special
conformal transformation and vector generators read
\begin{eqnarray}
D&=&tH-\frac12\int
dx\rho\sum_{k=0}^{2n}(-1)^{k}(k+1)\upsilon_i^{(k)}\upsilon_i^{(2n-k-1)},\nn
\\
K&=&t^2H-2tD-\frac12\int
dx\rho\sum_{k=0}^{2n}(-1)^k\Big((n+1)(2n+1)-k(k+1)\Big)\upsilon^{(k-1)}_i\upsilon^{(2n-k-1)}_i,\nn
\\
C_i^{(k)}&=&\sum_{s=0}^k(-1)^s\frac{k!}{(k-s)!}t^{k-s}\int dx \rho
\upsilon^{(2n-s)}_i,\quad  k=0,...,2n+1,
\end{eqnarray}
where $\upsilon^{(-1)}_i=x_i$. Under the Poisson brackets
(\ref{PBPFl}), the conserved charges obey the algebra (\ref{lCG}), which is
extended by the central charge \cite{GM11}
\begin{eqnarray}
\{C_i^{(k)},C_j^{(m)}\}&=&(-1)^{k}k!m!\delta_{(k+m)(2n+1)}\delta_{ij}M,\quad
M=\int dx\rho.
\end{eqnarray}

For a perfect fluid ($\ell=\frac12$, $n=0$) the Hamiltonian formulation involving non-canonical Poisson
brackets was originally given in \cite{MG80}.

\section{{Clebsch parametrization and Lagrangian formulation}}\label{Sec3}

In order to demonstrate how the equations (\ref{PFEqdl}) can be
obtained from the variational principle, let us first recall (for
more details see e.g. \cite{JNPP04}) how the Lagrangian for a
perfect fluid is built which correctly reproduces the continuity
equation and the Euler equation
\begin{eqnarray}\label{PFEq}
&&{\partial_0\rho}+ {\partial_i (\rho\upsilon_i)}=0\label{PFEqCon},\\
&&{\cal D}\upsilon_i=-\frac{1}{\rho}{\partial_i p}.\label{PFEqEul}
\end{eqnarray}

In three spatial dimensions this is achieved by making recourse to the Clebsch
parametrization for the velocity vector field
\begin{eqnarray}\label{PFCleb}
\upsilon_i=\partial_i\theta+\alpha\partial_i\beta,
\end{eqnarray}
which involves three scalar functions $\theta$, $\alpha$ and $\beta$. Then the
Lagrangian reads
\begin{eqnarray}
L&=&-\int
dx\rho\left(\partial_0\theta+\alpha\partial_0\beta\right)-H
\nn\\
&=&-\int
dx\rho\left(\partial_0\theta+\alpha\partial_0\beta\right)-\int
dx\left(\frac12\rho\upsilon_i\upsilon_i+V\right),
\end{eqnarray}
where $H$ is the Hamiltonian (the total energy) with $\upsilon_i$
in (\ref{PFCleb}). The variation under $\theta$ gives the
continuity equation (\ref{PFEqCon}), while the variations with respect to
$\alpha$ and $\beta$ give
\begin{eqnarray}\label{PFEqAdd}
{\cal D\alpha}=0,\quad {\cal D\beta}=0,
\end{eqnarray}
where (\ref{PFEqCon}) was taken into account.

Finally, varying with respect to $\rho$ and using (\ref{PFEqAdd}), one gets
\begin{eqnarray}
{\cal D}\theta-\frac12\upsilon_i\upsilon_i+V'_\rho=0.
\end{eqnarray}
As a result, the Euler equation (\ref{PFEqEul}) are
satisfied
\begin{eqnarray}\label{ConEq}
{\cal D}\upsilon_i={\cal
D}(\partial_i\theta+\alpha\partial_i\beta)=-\frac{1}{\rho}\frac{\partial
p}{\partial x_i},\quad p=\rho V'_\rho-V.
\end{eqnarray}

In order to generalize the construction above to the
$\ell$-conformal perfect fluid, we go over to the equivalent first
order system (\ref{HamFlEql}). In the case of
half-integer $\ell=n+\frac12$, the starting equations read
\begin{eqnarray}
&&{\partial_0\rho}+ {\partial_i
(\rho\upsilon^{(0)}_i)}=0\label{lPFEqLag1},\\
&& {\cal D}\upsilon^{(k)}_i=\upsilon^{(k+1)}_i,\quad k=0,1,...,2n-1\label{lPFEqLag2},\\
&& {\cal D}\upsilon^{(2n)}_i=-\frac{1}{\rho}{\partial_i p},\quad
p=\nu\rho^{1+\frac{1}{\ell d}}\label{lPFEqLag3}.
\end{eqnarray}
Note that these equations are completely equivalent to
(\ref{PFEqdl}) and hence completely characterize the
$\ell$-conformal perfect fluid.

A key ingredient of the construction above was the Clebsch
parametrization of the velocity vector variable. For the
$\ell$-conformal perfect fluid one has a set
of vector variables $\upsilon_i^{(k)}$ and it seems natural to expect that a
Clebsch-type decomposition will be needed for each of them. It
turns out, however, that in order to obtain the equations
(\ref{lPFEqLag1})-(\ref{lPFEqLag3}) from the variational principle
only the highest component $\upsilon_i^{(2n)}$
should be Clebsch-decomposed, while the remaining vector variables
$\upsilon_i ^{(k)}$ with $k<2n$ may remain intact. Up to a field redefinition, a suitable Clebsch-type
decomposition can be chosen in the form
\begin{eqnarray}\label{CleblPF}
\upsilon^{(2n)}_i=\partial_i\theta+\alpha\partial_i\beta
+\sum_{k=0}^{n-1}(-1)^{k+1}\upsilon_j^{(k)}\partial_i\upsilon_j^{(2n-k-1)}.
\end{eqnarray}
When $n=0$, there is no sum on the right hand side and the decomposition
for the Euler fluid (\ref{PFCleb}) is reproduced. The generalized
Lagrangian reads
\begin{eqnarray}\label{LaglPF}
L&=&-\int
dx\rho\left(\partial_0\theta+\alpha\partial_0\beta+\sum_{k=0}^{n-1}(-1)^{k+1}\upsilon_i^{(k)}\partial_0\upsilon_i^{(2n-k-1)}\right)
-H,
\end{eqnarray}
where $H$ is Hamiltonian (\ref{Hamdl}) with $\upsilon_i^{(2n)}$
in (\ref{CleblPF}). Thus, the basic variables for the
Lagrangian (\ref{LaglPF}) are the scalar fields $\rho$, $\theta$,
$\alpha$, $\beta$ and a set of vector fields $\upsilon_i^{(k)}$ with $k<2n$.

Let us demonstrate how the equations
(\ref{lPFEqLag1})-(\ref{lPFEqLag3}) follow from the Lagrangian
(\ref{LaglPF}). By varying the
Lagrangian with respect to $\theta$, one obtains the continuity
equation (\ref{lPFEqLag1}). Variations with respect to $\alpha$
and $\beta$ give (\ref{PFEqAdd}), as before. Varying
with respect to $\upsilon_i^{(k)}$ and taking into account
(\ref{lPFEqLag1}), the equations (\ref{lPFEqLag2}) are reproduced.
Finally, varying with respect to $\rho$ and using
(\ref{PFEqAdd}), one gets
\begin{eqnarray}
{\cal
D}\theta-\upsilon^{(0)}_i\upsilon^{(2n)}_i+\frac{(-1)^n}{2}\upsilon^{(n)}_i\upsilon^{(n)}_i+V'_\rho=0.
\end{eqnarray}
As a result, the equation
\begin{eqnarray}\label{LaglPFEq}
{\cal D}\upsilon^{(2n)}_i={\cal
D}\left(\partial_i\theta+\alpha\partial_i\beta+\sum_{k=0}^{n-1}(-1)^{k+1}\upsilon_j^{(k)}\partial_i\upsilon_j^{(2n-k-1)}\right)=-\frac{1}{\rho}{\partial_i
p},
\end{eqnarray}
is satisfied as well,
where $p=\rho V'_\rho-V$.

Because the Lagrangian (\ref{LaglPF}) involves only the first
temporal derivative, a transition to the Hamiltonian formalism
will lead to constraints. In the next section, we use
the Dirac method \cite{Dir64} to analyze such constraints
and demonstrate how the non-canonical Poisson brackets (\ref{PBPFl}) show up.

\section{Dirac's constraint analysis}\label{Sec4}

For simplicity of the presentation, let us focus on the $\ell=\frac32$ case. The corresponding
Lagrangian is given by (\ref{LaglPF}) with $n=1$
\begin{eqnarray}\label{Lag32PF}
L&=&-\int
dx\rho\left(\partial_0\theta+\alpha\partial_0\beta-\upsilon_i^{(0)}\partial_0\upsilon_i^{(1)}\right)-H\nn\\
&=&-\int
dx\rho\left(\partial_0\theta+\alpha\partial_0\beta-\upsilon_i^{(0)}\partial_0\upsilon_i^{(1)}\right)-\int
dx
\left(\rho\upsilon^{(0)}_i\upsilon^{(2)}_i-\frac12\rho\upsilon^{(1)}_i\upsilon^{(1)}_i+V\right),
\end{eqnarray}
where $H$ is the Hamiltonian (\ref{Hamdl}) with $\upsilon^{(2)}_i$
defined in (\ref{CleblPF})
\begin{eqnarray}\label{Cleb32}
\upsilon^{(2)}_i=\partial_i\theta+\alpha\partial_i\beta-\upsilon_j^{(0)}\partial_i\upsilon_j^{(1)}.
\end{eqnarray}
Further simplification occurs if one sets the scalar variables $\alpha$ and
$\beta$ to zero as particular solutions to the equations (\ref{PFEqAdd}).
This will not affect the final result but simplify the calculations.
In this case, the phase space
consists of basic variables
$X^A=(\rho,\theta,\upsilon_i^{(0)},\upsilon_i^{(1)})$ and their
conjugate momenta $P^A=(p_\rho,p_\theta,p_i^{(0)},p_i^{(1)})$, which obey the
canonical Poisson brackets
\begin{align}\label{CanonPB}
& \{\rho(x),p_\rho(y)\}=\delta(x-y), &&
\{\upsilon_i^{(0)}(x),p_j^{(0)}(y)=\delta_{ij}\delta(x-y),
\nn\\
& \{\theta(x),p_\theta(y)\}=\delta(x-y), &&
\{\upsilon_i^{(1)}(x),p_j^{(1)}(y)=\delta_{ij}\delta(x-y).
\end{align}
From the conditions determining the canonical momenta
 $P^A=\frac{\partial L}{\partial(\partial_0 X^A)}$ the following primary constraints arise
\begin{eqnarray}\label{Const}
\Phi^A\equiv\begin{pmatrix}
\phi_\rho\\
\phi_\theta\\
\phi_i^{(0)}\\
\phi_i^{(1)}\\
\end{pmatrix}=
\begin{pmatrix}
p_\rho\\
p_\theta+\rho\\
p_i^{(0)}\\
p_i^{(1)}-\rho\upsilon_i^{(0)}\\
\end{pmatrix}
=\begin{pmatrix}
0\\
0\\
0\\
0\\
\end{pmatrix}.
\end{eqnarray}
Then according to Dirac's method \cite{Dir64} the total
Hamiltonian reads
\begin{eqnarray}
H_T=H+\int
dx(\lambda_\rho\phi_\rho+\lambda_\theta\phi_\theta+\lambda_i^{(0)}\phi_i^{(0)}+\lambda_i^{(1)}\phi_i^{(1)}),
\end{eqnarray}
where $H$ is the canonical Hamiltonian in (\ref{Lag32PF}) and
$(\lambda_\rho,\lambda_\theta,\lambda_i^{(0)},\lambda_i^{(1)})$ are the
Lagrangian multipliers. Requiring the constraints to be conserved over
time, $\partial_0\phi=\{\phi,H_T\}=0$, one unambiguously specifies the Lagrangian multipliers
\begin{align}
&\lambda_\rho=-\partial_i(\rho\upsilon^{(0)}_i), &&
\lambda^{(0)}_i=\upsilon^{(1)}_i-\upsilon^{(0)}_j\partial_j\upsilon^{(0)}_i,
\nn\\
&
\lambda_\theta=\frac12\upsilon^{(1)}_i\upsilon^{(1)}_i+\upsilon^{(0)}_j(\upsilon^{(2)}_j-\partial_j\theta)-V'_\rho,
&&
\lambda^{(1)}_i=\upsilon^{(2)}_i-\upsilon^{(0)}_j\partial_j\upsilon^{(1)}_i.
\end{align}
The latter fact implies that all the constraints (\ref{Const}) are
second-class. The same conclusion is reached by analyzing the
Poisson brackets among the constraints
$\Phi_A=(\phi_\rho,\phi_\theta,\phi_i^{(0)},\phi_i^{(1)})$, which
form the non-degenerate matrix
\begin{eqnarray}
\Lambda_{AB}(x,x')=\{\Phi_A(x),\Phi_B(x')\}=
\begin{pmatrix}
0&-1&0&\upsilon_i^{(0)}\\
1&0&0&0\\
0&0&0&\rho\delta_{ij}\\
-\upsilon_i^{(0)}&0&-\rho\delta_{ij}&0
\end{pmatrix}_x\delta(x-x').
\end{eqnarray}
The inverse matrix reads
\begin{eqnarray}\label{ConstMatrInv}
\Lambda_{AB}^{-1}(x,x')=\{\Phi_A(x),\Phi_B(x')\}^{-1}=
\begin{pmatrix}
0&1&0& 0\\
-1&0&\frac{\upsilon_i^{(0)}}{\rho}&0\\
0&-\frac{\upsilon_i^{(0)}}{\rho}&0&-\frac{\delta_{ij}}{\rho}\\
0&0&\frac{\delta_{ij}}{\rho}&0
\end{pmatrix}_x\delta(x-x'),
\end{eqnarray}
such that $\int
dz\Lambda^{-1}_{AC}(x,z)\Lambda_{CB}(z,x')=\delta_{AB}\delta(x-x')$.

In order to make connection with the Hamiltonian formulation
presented in section \ref{Sec2}, one should resolve the constraints
(\ref{Const}) and deal with the Dirac bracket
\begin{eqnarray}\label{DB}
\{A(x),B(y)\}_D&=&\{A(x),B(y)\}+\int
dz\Big[\{A(x),\phi_\theta(z)\}\{\phi_\rho(z),B(y)\}\nn\\
&&-\Big(\{A(x),\phi_\rho(z)\}-\frac{\upsilon_i^{(0)}(z)}{\rho(z)}\{A(x),\phi_i^{(0)}(z)\}\Big)\{\phi_\theta(z),B(y)\}\nn\\
&&-\Big(\frac{\upsilon_i^{(0)}(z)}{\rho(z)}\{A(x),\phi_\theta(z)\}+\frac{1}{\rho(z)}\{A(x),\phi_i^{(1)}(z)\}\Big)\{\phi_i^{(0)}(z),B(y)\}\nn\\
&&+\frac{1}{\rho(z)}\{A(x),\phi_i^{(0)}(z)\}\{\phi_i^{(1)}(z),B(y)\}\Big],
\end{eqnarray}
where $A(t,x)$ and $B(t,x)$ are two arbitrary field variables of the
phase space.

By resolving the constraints, one eliminates the canonical momenta from
the consideration reducing the set of fields to $\rho$, $\theta$,
$\upsilon_i^{(0)}$ and $\upsilon_i^{(1)}$. Substituting them in
(\ref{DB}) and taking into account
(\ref{CanonPB}), one obtains the following non-zero Dirac brackets
\begin{eqnarray}\label{DB32}
&& \{\rho(x),\theta(y)\}_D=\delta(x-y),
\nn\\
&&\{\theta(x),\upsilon_i^{(0)}(y)\}_D=\frac{\upsilon_i^{(0)}}{\rho}\delta(x-y),
\nn\\
&&
\{\upsilon_i^{(0)}(x),\upsilon_j^{(1)}(y)\}_D=-\frac{1}{\rho}\delta_{ij}\delta(x-y).
\end{eqnarray}

When $\alpha$ and $\beta$ are not zero, it suffices to
add the following Dirac
brackets
\begin{eqnarray}\label{DB32Add}
\{\theta(x),\alpha(y)\}_D=\frac{\alpha}{\rho}\delta(x-y),\quad
\{\alpha(x),\beta(y)\}_D=\frac{1}{\rho}\delta(x-y).
\end{eqnarray}
Using (\ref{DB32}) and (\ref{DB32Add}), one can verify that the
non-canonical Poisson brackets (\ref{PBPFl}) are reproduced for
$n=1$ with $\upsilon_i^{(2)}$ defined in (\ref{Cleb32}). Also
one can easily identify the canonical pairs
$(\rho,\theta)$, $(\rho\alpha,\beta)$ and
$(\rho\upsilon_i^{(0)},\upsilon_i^{(1)})$. The same pairs result
from the Lagrangian (\ref{Lag32PF}).

The constraint analysis above can be readily generalized to
the case of arbitrary half-integer $\ell$. One can see from the Lagrangian (\ref{LaglPF})
that the canonical pairs include $(\rho,\theta)$,
$(\rho\alpha,\beta)$ and
$(\rho\upsilon_i^{(k)},\upsilon_i^{(2n-k-1)})$, where
$k=0,1,...,n-1$.

\section{Conclusion}\label{Sec5}

To summarize, in this work the Lagrangian formulation for the
generalized higher derivative perfect fluid equations, which hold
invariant under the $\ell$-conformal Galilei group with arbitrary
half-integer parameter $\ell$, was constructed. It is based on a
suitably chosen Clebsch-type parametrization and correctly
reproduces the Lagrangian description of a Euler fluid in
\cite{JNPP04} for $\ell=\frac12$. The Dirac method was used in order
to analyze constraints which arose after transition to the
Hamiltonian formalism. It was demonstrated that all the constraints
are second-class. The corresponding Dirac brackets were computed,
which reproduced the Hamiltonian description in \cite{Sne23a} given
in terms of non-canonical Poisson brackets.

The recent works on fluid mechanics with the $\ell$-conformal
Galilei symmetry was mostly focused on the development of the
mathematical structure. It now calls for physical applications.
Firstly, a clear-cut thermodynamic interpretation is needed. In this
regard, the approach in \cite{BHOSV} may pave the way. Possible link
to statistical mechanics, in particular the universality classes of
Hohenberg and Halperin \cite{HH77}, is interesting to explore.
Because the generalized perfect fluid equations contain higher
derivative terms, they may find potential applications within the
context of the hyperjerk theory \cite{CS06}.

Turning to other possible developments, it would be
interesting to develop the Lagrange picture \cite{AK88} for
describing higher derivative fluid mechanics and relate it to the
results presented in this paper. Supersymmetric extensions of the
Lagrangian (\ref{LaglPF}) in the spirit of \cite{JP00,Gal24} as well
as possible applications within the context of the fluid/gravity
correspondence are worth exploring.

\section*{Acknowledgements}
This work was supported by
the Russian Science Foundation, grant No 23-11-00002.

\end{document}